\documentclass[pra,twocolumn,english,showpacs]{revtex4}
\usepackage{graphicx}
\usepackage{amssymb}
\usepackage{times}

\begin{document}
\title{Thermometry of Fermionic Atoms in an Optical Lattice}
\author{Michael K\"ohl}
\affiliation{Institute of Quantum Electronics, ETH Z\"urich
H\"onggerberg, CH-8093 Z\"urich, Switzerland}
\begin{abstract}
Low temperatures are necessary for the observation of strongly
correlated quantum phases of fermionic atoms in optical lattices.
We analyze how the temperature of a Fermi gas is altered when the
fermions are loaded into an optical lattice with an underlying
harmonic confining potential and show how the temperature can be
measured. The temperature of the atoms in the optical lattice
determines the fraction of doubly occupied lattice sites of a
two-component Fermi gas. We analytically calculate this quantity
and find a strong temperature dependence. This fraction can be
measured by studying the production of molecules in the lattice
using a Feshbach resonance which allows for precise thermometry of
atoms in an optical lattice.
\end{abstract}
  \maketitle

Ultracold fermionic atoms in optical lattices are a promising tool
to simulate problems from condensed matter physics. Interesting
phase transitions between many-body quantum phases in the lattice
are predicted \cite{Hofstetter2002,Illuminati2004}. For these
phase transitions the temperature is one of the key control
parameters and a precise knowledge of the temperature of the atoms
in the lattice is necessary to experimentally determine the phase
diagram. Proposals for cooling the atoms in the lattice in order
to reach interesting quantum phases such as the BEC-BCS transition
in lattices or an anti-ferromagnetic phase have been devised
\cite{Hofstetter2002,Blakie2005,Werner2005}. However, standard
methods for thermometry -- such as observing the rounding-off of
the Fermi edge \cite{DeMarco1999} -- fail in the inhomogeneous
lattice since they turned out to be dominated by the trapping
potential rather than by temperature \cite{Rigol2004,Koehl2005}.

Previous theoretical investigations of ultracold fermions in
optical lattices have often neglected the role of the harmonic
confining potential. For ongoing experiments, this confining
potential is an unavoidable ingredient and its consequences, such
as an inhomogeneous filling of the lattice and the co-existence of
insulating and conducting regions, must be considered. In a
typical experimental sequence a quantum degenerate Fermi gas is
initially prepared in a purely harmonic optical dipole potential
and subsequently transferred into a three-dimensional optical
lattice with underlying confining potential \cite{Koehl2005}.

The transfer of a Fermi gas from a harmonic trapping potential
into an optical lattice leads to a significant modification of the
density of states. A change in the density of states can result in
both an adiabatic cooling or an adiabatic heating of the gas. We
find that for the situation realized in the experiments the gas is
adiabatically heated which poses a difficulty for future
experiments aiming at very low temperatures. To determine the
temperature in the lattice we analyze the fraction of doubly
occupied sites of a noninteracting two-component Fermi gas
\cite{Katzgraber2005}. This can be measured experimentally by
studying the formation of dimers in the optical lattice on a
Feshbach resonance \cite{Stoferle2005} or using photo association
\cite{Rom2004,Xu2005,Ryu2005}. We calculate the fraction of doubly
occupied sites analytically and find a strong temperature
dependence which makes it ideal for thermometry of ultracold
fermions in an optical lattice.

%Model
A noninteracting two-component Fermi gas in an optical lattice
with an underlying harmonic confining potential can be described
by the Hamiltonian
\begin{equation}
H_\sigma=-t\sum_{<j,l>} c^\dag_{j\sigma} c_{l\sigma} + \frac{m}{2}
\omega^2 d^2\sum_j j^2 n_{j\sigma}. \label{hamiltonian}
\end{equation}
Here $\sigma=\{\uparrow,\downarrow\}$ denotes the spin of the
particle, $t$ denotes the hopping matrix element between adjacent
lattice sites, $c^\dag_{j\sigma}$ and $c_{j\sigma}$ are the
creation and the annihilation operators on the lattice site $j$,
respectively, and $n_{j\sigma}=c^\dag_{j\sigma}c_{j\sigma}$ is the
number operator. Furthermore,  $m$ denotes the atomic mass,
$\omega$ the trapping frequency of the harmonic confining
potential and $d=\lambda/2$ the lattice spacing given by the laser
wavelength $\lambda$. The interactions between atoms in different
spin states are assumed to be zero, as realized experimentally by
using a Feshbach resonance \cite{Koehl2005}. The trapping
frequency $\omega$ of the confining potential is determined by the
potential depth $V$ and the waist $w$ of the Gaussian intensity
profile of the laser beam. For a three-dimensional lattice
generated by three mutually orthogonal laser beams of identical
waist $w$ and potential depth $V$ one obtains
$\frac{m}{2}\omega^2=\frac{4 V}{w^2}$.

Let us first consider the single particle density of states
\cite{Hooley2004} in the low-tunnelling limit in a
three-dimensional optical lattice. There the particles are
localized in the potential wells and we restrict ourselves to a
single band approximation which is justified by the experiment
\cite{Koehl2005}. For determining the density of states we assume
the width of the lowest Bloch band $4t$ to be smaller than the
energy offset between neighboring lattice sites due to the
harmonic confinement. This approximation is valid except for a
very small region at the center of the harmonically trapped cloud,
where delocalized atoms should be considered. However, only few
atoms are affected and we neglect their contribution in the
analytical calculation. The presence of the harmonic confinement
modifies the density of states significantly: for a particle on
the lattice site $j$, the potential energy is $\frac{m}{2}\omega^2
\left(\frac{\lambda}{2}\right)^2 j^2$. Allowing for only one
particle per lattice site, the density of states in three
dimensions reads
\begin{equation}
\rho_{3D}(E)=2 \pi E^{1/2}\left(\textstyle{\frac{m
\omega^2\lambda^2}{8}} \right)^{-3/2}. \label{rho3d}
\end{equation}
This shows that the density of states of fermions localized in an
optical lattice with underlying harmonic confinement has the same
power law dependence as that of {\it free} fermions. This analogy
can be understood by the following argument: in free space the
fermions are characterized by their momentum $k$ and the
eigenstates are equally spaced in momentum space. Each fermion has
a kinetic energy $E\propto k^2$. In the low-tunnelling limit with
harmonic confinement the fermions occupy equally spaced lattice
sites at positions $j\lambda/2$ and have a potential energy
$E\propto j^2 (\lambda/2)^2$ only, which results in the same
dependence of the density of states on $E$ as in the free fermion
case. From the density of states we directly obtain the Fermi
energy in the tight-binding limit:
\begin{equation}
E_F(N)=\frac{m \omega^2 \lambda^2}{8} \left(\frac{3N}{4
\pi}\right)^{2/3}.
\end{equation}
In one and two dimensions we find the same analogy to the free
fermion case, namely $\rho_{2D}(E)=8 \pi/(m \omega^2 \lambda^2)$
and $\rho_{1D}(E)=( m \omega^2 \lambda^2 E/8)^{-1/2}$, or
generally $\rho(E) \propto E^{d/2-1}$ with $d$ being the
dimension.

%\begin{figure}[ht]
%\begin{center}
%  \includegraphics[width=.95\columnwidth]{fig_1.eps}
%\end{center}
%  \caption{The exponent of the density of states for various lattice parameters obtained from a numerical calculation.
%  The solid line corresponds to a waist of $w=70\,\mu$m, the dotted line $w=90\,\mu$m, and the dashed line $w=50\,\mu$m.
%  {\bf a)} In the three-dimensional configuration the exponent approaches 1/2. {\bf b)} In the two-dimensional configuration
%  the exponent approaches  the value 0 asymptotically. $E_r=h^2/2 m \lambda^2$ is the recoil energy.}
%  \label{fig1}
%\end{figure}

To estimate the validity of the low-tunnelling approximation we
compare the analytical results with a numerical calculation where
we diagonalize the Hamiltonian (\ref{hamiltonian}) exactly. We
compute the density of states and fit the result with a function
$\widetilde{\rho}(E)=a\cdot E^\nu$ to obtain the exponent $\nu$.
For a large potential depth, the value of $\nu$ approaches 1/2, as
expected from equation (\ref{rho3d}). Similarly, for a
two-dimensional quantum gas in an optical lattice the exponent
approaches 0. The calculations were performed on $300^2$ lattice
sites in 2D and $100^3$ lattice sites in 3D.

When atoms are loaded into an optical lattice the density of
states changes from the three-dimensional harmonic oscillator
result $\rho(E) \propto E^2$ to the low-tunneling results
$\rho(E)\propto E^{d/2-1}$. Adiabatically changing the density of
states of a system can be used to cool or heat a gas, provided
that collisions keep the gas in thermal equilibrium
\cite{Pinkse1997,StamperKurn1998}. We assume an adiabatic
evolution of the gas during the loading procedure which implies
that no heat is exchanged with the environment. The interactions
between the atoms are supposed to be sufficiently large to ensure
thermalization of the sample but small enough so that the energy
spectrum of the gas is not altered. In such a situation the gas
evolves along a path of constant entropy $S$. The entropy $S$ of a
Fermi gas is given by \cite{Huang1987}
\begin{equation}
\frac{S}{k_B}= \frac{\mathcal{E}-\mu N}{k_B
T}+\sum_n\log(1+e^{(\mu-E_n)/k_BT}) \label{entropy}
\end{equation}
with the total energy $\mathcal{E}=\int_{-\infty}^\infty E \rho(E)
f(E) dE$, the chemical potential $\mu$ and the temperature $T$.
$n$ labels the energy eigenstates $E_n$ and
$f(E)=\left(e^{(E-\mu)/k_BT}+1\right)^{-1}$ is the Fermi
distribution function. The chemical potential is determined from
the normalization to the total particle number
$N=\int_{-\infty}^\infty \rho(E) f(E) dE$. For low temperatures
the entropy in the two limiting cases can be obtained using the
Sommerfeld approximation. For noninteracting fermions trapped in a
three-dimensional power-law potential $V({\bf r})\propto
\sum\limits_{m=x,y,z}r_m^\alpha$ the entropy is given by
\cite{Su2003,Carr2004}
\begin{equation}
\frac{S}{k_B}=N
\pi^2\left(\frac{1}{\alpha}+\frac{1}{2}\right)\frac{k_BT}{E_F}
+\mathcal{O}\left((\textstyle{\frac{k_BT}{E_F}})^2\right).
\label{entropy2}
\end{equation}
We now compare the result of the entropy for a given number of
particles in a harmonic trap ($\alpha=2$) with an ideal Fermi gas
in an optical lattice with underlying harmonic confinement. As
discussed above, the latter has the same density of states as free
fermions in a box potential and therefore corresponds to
$\alpha=\infty$. Consequently, the entropy in a harmonic trap is a
factor of $2$ larger than for the fermions in the optical lattice.
Thus when the transfer from the harmonic potential into the
lattice is performed adiabatically it is accompanied by a
temperature increase by a factor 2. The effect of the turn-on of
the lattice on the atom number statistics at the lattice sites has
been calculated for a one-dimensional Fermi gas \cite{Budde2004}.

\begin{figure}[ht]
  \includegraphics[width=\columnwidth]{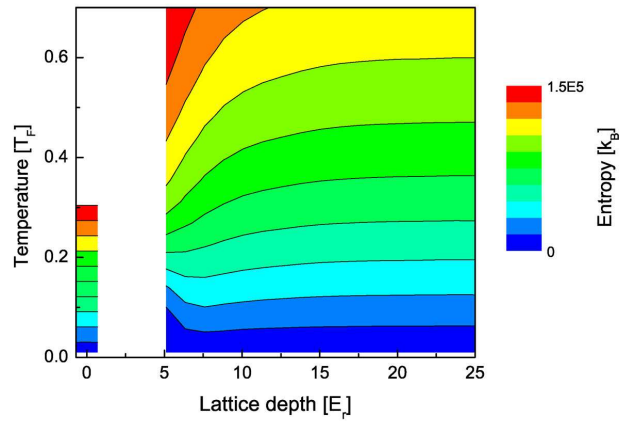}
  \caption{Curves of constant entropy in the lattice. The calculation is performed by diagonalizing Hamiltonian (\ref{hamiltonian}) with
  $N=50000$ atoms trapped in three-dimensional optical lattice with the waist of the lattice laser being $w=70\,\mu$m, according to
  the situation in \cite{Stoferle2005}. The values shown at zero lattice depth are the harmonic oscillator result without lattice.
  For $s=V/E_r\lesssim 5$ the tight-binding Hamiltonian may give incorrect results and therefore we refrain from numerical calculations in this regime.}
  \label{fig2}
\end{figure}

We numerically calculate $S$ for various temperatures and lattice
depths (see fig. \ref{fig2}), similar to what has been done
previously for the homogeneous lattice \cite{Blakie2005}. For a
large potential depth the numerical calculations agree with the
analytical prediction of equation (\ref{entropy2}). Moreover, when
moving along a line of constant entropy as the lattice depth is
increased we observe that the temperature never drops below its
initial value in the harmonic oscillator trap. This implies that
no cooling can be achieved during the transfer into the lattice.

The quest for obtaining ultra-low temperatures in the optical
lattice is accompanied with the need for precise thermometry. We
show that the temperature can be obtained from the fraction of
doubly occupied sites of a noninteracting two-component Fermi gas.
A similar problem has been studied numerically for bosonic
\cite{Pupillo2004} and fermionic \cite{Katzgraber2005} atoms. We
analytically calculate the number of doubly occupied sites $N_2$
assuming that both species are distributed according to the same
Fermi distribution function $f(E)=f_\uparrow(E)=f_\downarrow(E)$
\begin{equation}
N_2=\int_{-\infty}^\infty \rho(E) f^2(E) dE \label{eq1}
\end{equation}

In two dimensions (i.e. for $\nu=0$) simple analytic expressions
are obtained. We consider a Fermi gas strongly confined along one
axis to create a two-dimensional quantum gas which is subject to
two crossed optical lattices in the transverse directions. The
chemical potential as a function of temperature is $\mu=k_BT
\ln\left(e^{E_F/k_BT}-1\right)$ and the fraction of doubly
occupied sites is
\begin{equation}
n_{2}= \frac{N_2}{N}=\frac{k_B T}{E_F} \left(
e^{-E_F/k_BT}+\frac{E_F}{k_BT}-1\right).\label{prediction2D}
\end{equation}

\begin{figure}[ht]
  \includegraphics[width=\columnwidth]{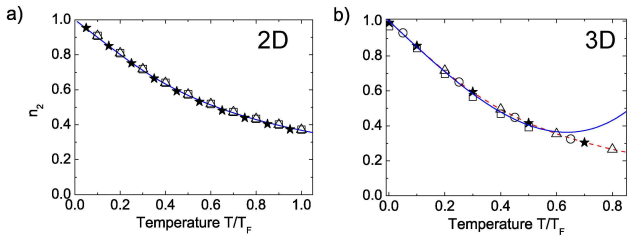}
  \caption{{\bf a)} Comparison of the analytical formula (\ref{prediction2D}) (solid line)
  with the numerically calculated double occupation of a two-dimensional quantum gas in an optical lattice
  for various parameters by diagonalizing Hamiltonian (\ref{hamiltonian}). Squares: $s$=25, $w$=70\,$\mu$m, $N$=5000, stars: $s$=15, $w$=100\,$\mu$m, $N$=3500,
  triangles: $s$=15, $w$=50\,$\mu$m, $N$=2500.
  {\bf b)} Comparison of the analytical formula (\ref{prediction3D}) (solid line) with the numerically calculated occupation in a 3D optical lattice
  for various experimentally relevant parameters by diagonalizing Hamiltonian (\ref{hamiltonian}). Circles: $s$=15, $w$=50\,$\mu$m, $N$=25000, triangles: $s$=20, $w$=70\,$\mu$m, $N$=15000,
  squares: $s$=15, $w$=100\,$\mu$m, $N$=40000, stars: $s$=25, $w$=130\,$\mu$m, $N$=25000, triangles: $s$=20, $w$=70\,$\mu$m, $N$=15000.
  The dashed line is a numerical solution of equation (\ref{eq1}) for the 3D lattice using the density of states given by eq. (\ref{rho3d}).}
  \label{fig3}
\end{figure}

For the general case, we use a low-temperature expansion similar
to the Sommerfeld expansion for free electrons. We integrate
eq.\,(\ref{eq1}) by parts and expand $R(E)=\int \rho(E) dE$ in a
power series around $E=\mu$, analogous to the method described in
\cite{Ashcroft1976}. We obtain the number of doubly occupied
lattice sites to be
\begin{equation}
N_2=\int_{-\infty}^\mu \rho(E) dE+ \sum_{n=1}^\infty a_n (k_B T)^n
\frac{d^{n-1}\rho(\epsilon)}{d \epsilon^{n-1}} |_{\epsilon=\mu}
\label{expansion3d}
\end{equation}
with the coefficients $a_n$ defined by
\begin{equation}
a_n=\int_{-\infty}^{\infty}\frac{x^n}{n!}\left(-\frac{d}{dx}\frac{1}{(e^x+1)^2}
\right) dx
\end{equation}
and given by $a_n=(2-2^{2-n})\zeta(n)$
for even values of $n$ and $a_{n}=-a_{n-1}$ for odd values of $n$.
The Sommerfeld expansion yields $N=\int_{-\infty}^\mu
\rho(E)dE+\sum_{n=1}^\infty b_{2n} (k_B T)^{2n}
\frac{d^{2n-1}\rho(\epsilon)}{d\epsilon^{2n-1}}|_{\epsilon=\mu}$
\cite{Ashcroft1976} with the coefficients
$b_{2n}=\int_{-\infty}^{\infty}\frac{x^{2n}}{(2n)!}\left(-\frac{d}{dx}\frac{1}{e^x+1}
\right) dx=a_{2n}$, which represent the even terms in the
expansion (\ref{expansion3d}). The analytic expression for the
fraction of doubly occupied lattice sites is then given by
\begin{eqnarray}
n_2=\frac{N_2}{N}=1+\frac{1}{N}\sum\limits_{n=0}^\infty a_{2n+1}
(k_BT)^{2n+1}
\frac{d^{2n}\rho(\epsilon)}{d\epsilon^{2n}}|_{\epsilon=\mu}.
\label{n2general}
\end{eqnarray}
For a density of states $\rho \propto E^\nu$ the fraction of
doubly occupied sites is given to first order in $k_BT/E_F$  by
\begin{equation}
n_2=1-(\nu+1)\frac{k_BT}{E_F}+\mathcal{O}\left((\textstyle{\frac{k_BT}{E_F}})^2\right).
\end{equation}
This expression shows that there is a linear dependence of the
fraction of doubly occupied sites on temperature in the degenerate
regime $k_B T \ll E_F$. The slope depends only on the exponent of
the density of states, but not on lattice depth (see figure
\ref{fig3}). This makes this quantity ideally suited for
thermometry.

For the special case of a deep three-dimensional optical lattice
with $\nu=1/2$ (see equation (\ref{rho3d})) we use the known
expression for the chemical potential
$\mu=E_F\left(1-\frac{\pi^2}{12}(\frac{k_BT}{E_F})^2\right)$ and
expand $n_2$ to third order in $k_BT/E_F$:
\begin{equation}
n_{2}=1-\frac{3}{2}\frac{k_B
T}{E_F}+\frac{\pi^2}{8}\left(\frac{k_BT}{E_F}\right)^3+\mathcal{O}\left((\textstyle{\frac{k_BT}{E_F}})^4\right).
\label{prediction3D}
\end{equation}

\begin{figure}[ht]
  \includegraphics[width=\columnwidth]{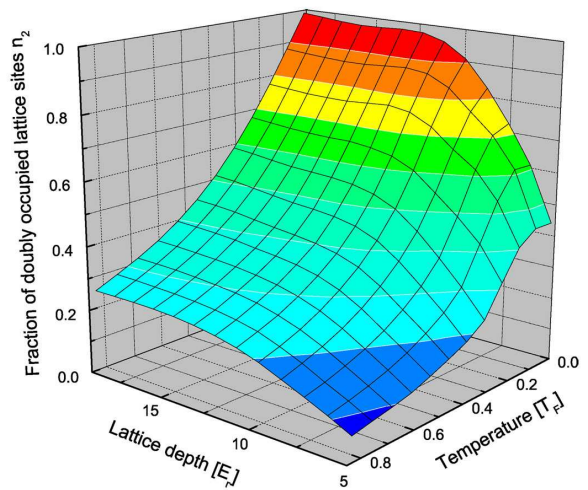}
  \caption{Fraction of doubly occupied lattice sites $n_2$ in a three-dimensional optical lattice as a function of the lattice depth and
  the temperature obtained by diagonalizing Hamiltonian (\ref{hamiltonian}). The particle number is $N=15000$ per spin state and the
  waist of the lattice laser is $w=70\,\mu$m, according to
  the situation in \cite{Koehl2005}.}
  \label{fig4}
\end{figure}

The dependence of the fraction of doubly occupied sites on
temperature in three dimensions is stronger than in two dimensions
but in both cases we find for $k_BT\ll E_F$ a predominantly linear
dependence with a slope of order unity. For comparison we compute
 the fraction of doubly occupied lattice sites numerically. In
figure \ref{fig3}a we compare the results in two dimensions with
the exact analytical prediction of equation (\ref{prediction2D}),
which shows excellent agreement. In figure \ref{fig3}b the
numerical results for a three-dimensional optical lattice are
compared with the expression (\ref{prediction3D}), which is valid
for $k_BT\ll E_F$. We also find very good agreement in this
situation. These results are also in agreement with the numerical
results of \cite{Katzgraber2005}.

When the system is not in the low-tunnelling regime we leave the
range of validity of the analytic calculation. To quantify this
limit, we have numerically studied how the fraction of doubly
occupied lattice sites varies with the potential depth of the
lattice (Fig. \ref{fig4}). We have chosen the experimentally
realized values of $N=15000$ per spin state and a waist of
$w=70\,\mu$m \cite{Koehl2005} as parameters. From the calculations
we observe that the analytical solution, which is independent of
the lattice depth, is good for a potential depth $V\gtrsim
12\,E_r$. In this low-tunnelling regime, $n_2$ depends on
temperature only. For smaller values of the lattice depth
numerical calculations are necessary to relate a measured fraction
of doubly occupied lattice sites to the temperature.

Let us now compare our results with a recent experiment
\cite{Stoferle2005}. A noninteracting, two-component Fermi gas
with an initial temperature in the harmonic trap of $T/T_F\simeq
0.25$ is loaded into a three-dimensional optical lattice. In a
deep optical lattice ($V=15\,E_r$) molecules are created by
adiabatically sweeping a magnetic field across a Fesh\-bach
resonance and the molecule fraction is determined to be
$n_2=0.43$. From our theory we calculate a temperature of
$T/T_F=0.46$ in the lattice. Provided that the sample in the
lattice is in thermal equilibrium (which, however, has not been
verified experimentally), this indicates a temperature increase of
approximately a factor two as compared to the initial situation.
This is expected for adiabatic loading of the optical lattice due
to entropy conservation in the loading process.

Finally, we would like to address the weakly interacting
two-component Fermi gas in an optical lattice. We take the on-site
interaction between different spin states of strength $U$ into
account using a mean-field approach \cite{Carr2004}, which
requires that $k_F |a|\ll 1$. The occupation number
$\nu_\sigma(r_i)$ per spin state $\sigma$ at a lattice site $r_i$
in the low-tunnelling regime is given by
$\nu_\sigma(r_i)=(\exp[(V(r_i)+U\nu_\uparrow(r_i)\nu_\downarrow(r_i)/2-\mu)/k_BT]+1)^{-1}$.
This equation is solved self-consistently. The results show that
the fraction of doubly occupied sites changes with the strength of
$U$: For repulsive interaction $n_2=\sum_i
\nu_\uparrow(r_i)\nu_\downarrow(r_i)/N$ is diminished and for
attractive interaction $n_2$ is increased. However, the
interaction induced effect is smaller than the temperature effect.
For $k_F|a|=0.05$, $w=70\,\mu$m, $N=15000$ (the parameters of
figure \ref{fig4}) and a lattice depth of 25\,$E_r$ we observe a
change of double occupancy by approximately $2\%$, which is small
as compared to the effect of temperature.

In conclusion, we have investigated a Fermi gas in an optical
lattice with an underlying harmonic trapping potential. Our
results have immediate consequences for ongoing experiments with
ultracold fermions in optical lattices. We find that the Fermi gas
is adiabatically heated by approximately a factor two when the gas
is loaded into the lattice. Moreover, we calculate how the
occupation of an optical lattice depends on temperature and find
that the fraction of doubly occupied lattice sites for a
noninteracting, two-component Fermi gas is a sensitive quantity
for thermometry in the lattice.

We would like to thank T. Bourdel, C. Chin, T.\,Esslinger,  K.
G{\"u}nter, H. G. Katzgraber,  C. Kollath, H.\,Moritz, and
T.\,St{\"o}ferle for helpful discussions.

\end{document}